\documentstyle[prl,aps,twocolumn]{revtex}
\title{Electron correlation resonances in the transport through a single
quantum level}
\author{A. Levy Yeyati, A. Mart\'{\i}n-Rodero, and F. Flores}
\address{
Departamento de F\'\i sica de la Materia Condensada C-XII.\\
Facultad de Ciencias. Universidad Aut\'onoma de Madrid.\\
E-28049 Madrid. Spain.}
\begin{document}

\draft
\maketitle

\begin{abstract}
Correlation effects in the transport properties of a single quantum
level coupled to electron reservoirs are discussed theoretically using
a non-equilibrium Green functions approach. Our method is based on the
introduction of a second-order self-energy associated with
the Coulomb interaction that consistently eliminates the pathologies
found in previous perturbative calculations. We present results for the
current-voltage characteristic illustrating the different
correlation effects that may be found in this system, including the
Kondo anomaly and Coulomb blockade. We finally discuss the experimental
conditions for the simultaneous observation of these effects in an
ultrasmall quantum dot.
\end{abstract}
\vspace{0.1in}

\narrowtext

The transport properties of a quantum dot coupled to electron
reservoirs have received considerable attention in recent years both
from the experimental \cite{Exp} and the theoretical
\cite{Theo} sides. With the advent of nanotechnologies,
an ideal experimental device where tunneling takes place through a
single electronic level within the dot is now feasible. Different
theoretical papers \cite{Hersh,Meir93,Ng,Groshev} have suggested that the
tunneling through this single quantum level can be appropriately
described by the Anderson hamiltonian

\begin{eqnarray}
H & = & \sum_{\sigma} \epsilon_0 n_{0 \sigma}
+ \sum_{\nu,k,\sigma}
T^{\nu}_{k} (c^{\nu \dagger}_{k \sigma} c_{0 \sigma} +
c^{\dagger}_{o \sigma} c^{\nu}_{k \sigma}) \nonumber \\
& & + \sum_{\nu,k,\sigma}
(\epsilon^{\nu}_{k} + \mu^{\nu}) n^{\nu}_{k \sigma}
+ U n_{0 \uparrow} n_{0 \downarrow},
\end{eqnarray}

\noindent
where $\epsilon_0$ represents the quantum dot level, $U$ is the
effective Coulomb repulsion within the dot, $\epsilon^{\nu}_k + \mu^{\nu}$
with $\nu =$ Left or Right denotes the reservoir single-particle energies,
$\mu^{L} - \mu^{R} = eV$ being the applied bias and $T^{\nu}_k$ the
coupling between the reservoir states and the quantum dot level.

Hamiltonian (1) is expected to describe the main correlation effects
associated with the quantum dot. In particular, a Kondo-like resonance
near the Fermi energy should be reflected in the current-voltage
characteristic at low voltages, when the dot level is nearly
half-filled. On the other hand, a Coulomb blockade effect should lead to
a reduction in the conductance up to voltages large enough to overcome
the Coulomb repulsion in the dot level. Previous theoretical efforts have
mainly been concerned with
one of these two effects. In references \cite{Ng&Lee,Hersh,Meir93,Ng}
the Kondo-effect is studied using a variety of techniques both in the
linear and non-linear regimes, whereas the charging effects has been
analyzed in references \cite{Groshev,Chen&Ting}.
The aim of this paper is to present a solution for this model
which would allow us to describe accurately all the physically relevant
regimes, showing how many-body
resonances appear in the current-voltage characteristic.
We pay particular attention to the situation where features related to
the Kondo and charging effects may be simultaneously found.
Based on these
results, we shall also discuss in detail the experimental conditions
necessary to observe those resonances in semiconductor
nanostructures.

The transport properties of this model can be analyzed by using the
Keldysh formalism \cite{Keldysh}, where the retarded, $G^r$, and the
distribution, $G^{+-}$, Green functions are defined as follows

\begin{eqnarray}
G_{i,j}^{r}(w) & = & -{\it i} \int  \theta(t) < c^\dagger_j(t)
c_i(0) + c_i(0) c_j^{\dagger}(t) >   e^{{\it i} wt} dt
\nonumber \\
G_{i,j}^{+-}(w) & = & {\it i} \int < c^\dagger_j(t)
c_i(0) >   e^{{\it i} wt} dt .
\end{eqnarray}

These Green functions can be calculated by starting with the $L$ and $R$
reservoirs decoupled from the dot level. This case defines the
unperturbed Green functions ${\bf G}^{(0)}(w)$. Then ${\bf G}(w)$ can
be obtained by coupling
the dot level to the reservoirs and by introducing the self-energies
${\bf \Sigma}^r$ and ${\bf \Sigma}^{+-}$ which take into account the
Coulomb correlations  within the dot.
%

Once the different Green functions are obtained from the
corresponding Dyson equations \cite{Keldysh}, the current intensity
$I_{\nu}$ between reservoir ${\nu}$ and the dot level, is given by

\begin{eqnarray}
I_{\nu} &=& \frac{2e}{\hbar} \sum_k T^{\nu}_k \int dw \left[
G^{\nu +-}_{k,0} (w) - G^{\nu +-}_{0,k} (w) \right] .
\end{eqnarray}

The crucial point in order to solve this problem is to find a reasonable
approximation for the self-energies. For $U = 0$, ${\bf \Sigma} = 0$ and
the exact ${\bf G}(w)$ can be easily obtained using conventional Green
function techniques \cite{Caroli}.

The effect of a finite $U$ can be included by using perturbation theory
in $U$. This perturbative approach has been extensively analyzed by
Yamada and Yoshida \cite{YY} and Zlatic
and Horvatic \cite{ZH} (hereafter referred to as YY and ZH)
for the equilibrium Anderson Model (zero applied bias).
Hershfield et al. \cite{Hersh} have extended this approach to the
nonequilibrium
case by calculating $\Sigma^r_{0\sigma}(w)$ and $\Sigma^{+-}_{0\sigma}(w)$
up to second order in $U$.

In YY and ZH approach,
$\Sigma^{(2)}_{0 \sigma}(w)$ is calculated from the second-order
diagram
shown in figure 1 (inset), where each Green function line is a Hartree
Fock dressed propagator, whose retarded part is given by:

\begin{eqnarray}
G^{r HF}_{0 \sigma}(w) &=& \frac{1}{w - \bar{\epsilon}_{0 \sigma}
+ i \Gamma_L(w)  + i \Gamma_R(w)} ,
\end{eqnarray}

\noindent
where
$\bar{\epsilon}_{0 \sigma} = \epsilon_0 + U <n_{0 \bar{\sigma}}>$
and $\Gamma_\nu = 2\pi\sum_k \mid T_k^\nu \mid^2\delta(w-\epsilon_k)$.

Although this approximation gives a good description of the electron
correlation effects in the symmetric case ($\epsilon_0 = -U/2$), it
presents some drawbacks when one moves away from this condition.
In the equilibrium case this is clearly illustrated by the failure
of the Friedel-Langreth \cite{FL} (hereafter referred as FL) sum rule,
that relates the ``impurity" charge $<n_{0 \sigma}>$
%
%
to the phase shift created by the impurity at the Fermi energy
$ \eta(0) = Im \;\left[ \; ln \; G^r_{0 \sigma} (0)\; \right] $.

Figure 1a shows the degree of fulfillment of the FL
sum rule, $<n_{0 \sigma}> = -\frac{1}{\pi} \eta(0)$, using the second
order self-energy $\Sigma^{r (2)}_{0 \sigma}(w)$ in the YY-ZH
approach. In this figure we
compare $<n_{0 \sigma}>$ with $-\frac{1}{\pi}\eta(0)$ as a function of
the dot level $\epsilon_0$, for $U/\Gamma = 2.4\pi$,
neglecting the frequency dependence in $\Gamma_\nu$, and taking
$\Gamma_L = \Gamma_R = \Gamma/2$.

Improvements over the above $\Sigma^{(2)}_{0 \sigma}$ are not easy to obtain.
For instance, one could insert the full self-consistent dressed propagators
instead of the Hartree-Fock ones for calculating the diagram in figure 1.
This sort of self-consistent perturbation
theory is charge conserving \cite{Baym} and would verify the FL sum rule.
However, it has been shown that it
leads to a poorer description of the quasiparticle spectral-density
\cite{White}.

Our proposal to improve the YY and ZH approach is the following:
instead of using the Hartree-Fock solution as the initial one-electron
problem for the calculation of $\Sigma_{0\sigma}^{(2)}$, we
use a different self-consistent field as a starting point. In this initial
one-electron problem the Hartree-Fock level, $\bar{\epsilon}_{0 \sigma}$,
in Eq.(4) is replaced by an
effective level $\epsilon_{eff \sigma}$, which is determined by imposing
self-consistency between the initial and the final dot level occupancies.
As shown in figure 1b, this way of calculating the second order self-energy
gives a solution that closely verifies the FL sum rule.

Notice that this
procedure is formally correct within perturbation theory: we have
decomposed $H = H_{eff} + V$, where $H_{eff}$ is the one-electron part
of hamiltonian (1) with $\epsilon_0$ replaced by $\epsilon_{eff \sigma}$,
and we treat $V = H - H_{eff}$ as a perturbation.
The fulfillment of the FL sum rule and the condition of charge consistency
provides an appealing interpretation for $\epsilon_{eff \sigma}$: it must
behave like $\epsilon_0 + Re \Sigma^r_{0 \sigma}(0)$, the effective
potential at the Fermi energy. Obviously, for the symmetric case
our approach coincides with that of YY and ZH.

This procedure allows us to get a good description of the Kondo-like
peak  appearing around the Fermi energy and the overall behavior of the
spectral density for a broad range of parameters ($\epsilon_0$,
$U/\Gamma$). However, for high values of $U/\Gamma$ and far from the
symmetric situation, the position and the spectral weights of the
resonances
around $\epsilon_0$ and $\epsilon_0 + U$ are not so well described
\cite{Alv}. The reason is that $\Sigma^{r (2)}_{0 \sigma}$ does not
yield the atomic limit when $w \sim U \gg \Gamma$.
To improve the solution given above in this region we
introduce the following self-energy \cite{Alv}

\begin{equation}
\Sigma^r_{0 \sigma} (w) = \frac{\Sigma^{r (2)}_{0 \sigma}(w)}{1 - \alpha
\Sigma^{r (2)}_{0 \sigma} (w)} ,
\end{equation}
\noindent
with

\[ \alpha = \frac{(1- <n_{0 \bar{\sigma}}>)U + \epsilon_0 -
\epsilon_{eff}}{<n_{0 \bar{\sigma}}>(1 - <n_{0 \bar{\sigma}}>)U^2} ,
\]

\noindent
which has the virtue that it yields the appropriate
atomic limit \cite{Alv,Schrieffer}, for $w \sim U$ and $\Gamma/U
\rightarrow 0$
and behaves like $\Sigma^{r (2)}_{0 \sigma}$ for $U/\Gamma
\rightarrow 0$.

Let us now turn our attention to the non-equilibrium situation.
For the sake of simplicity, we shall
assume that in Eq. (1) $\mu^L = -\mu^R = eV/2$.
Following our approach for the equilibrium Anderson model,
we obtain the different nonequilibrium self-energies, $\Sigma_{0
\sigma}^r$ and $\Sigma_{0 \sigma}^{+-}$, by
introducing the effective levels $\epsilon_{eff \sigma}$, $\mu_{eff}^L$
and $\mu_{eff}^R$ in the initial one-electron hamiltonian

\begin{eqnarray}
H_{eff}^{(transport)} & = & \sum_{\sigma} \epsilon_{eff \sigma} n_{0 \sigma} +
\sum_{\nu,k,\sigma} (\epsilon^{\nu}_{k} + \mu^{\nu}_{eff})
n^{\nu}_{k \sigma}   \nonumber\\
& & + \sum_{\nu,k,\sigma}
T^{\nu}_{k} (c^{\nu \dagger}_{k \sigma} c_{0 \sigma} +
c^{\dagger}_{0 \sigma} c^{\nu}_{k \sigma} ) .
\end{eqnarray}

As a natural extension of our procedure for the equilibrium case,
$\epsilon_{eff \sigma}$, $\mu_{eff}^L$ and $\mu_{eff}^R$  are
determined by imposing self-consistency in the dot level charge
and in the currents $I_L$ and $I_R$ defined by Eq. (3).

Eq. (6)
allows us to calculate the one electron Green functions ${\bf G}^{eff}$
that will be used to obtain $\Sigma^{r(2)}_{0 \sigma}$ and
$\Sigma^{+-(2)}_{0 \sigma}$
by means of a second
order perturbative calculation, and finally
$\Sigma^r_{0\sigma}(w)$ and $\Sigma^{+-}_{0\sigma}(w)$ by using the
equivalent of Eq. (5) for the non-equilibrium case.

Let us remark that our method eliminates, in a natural way, the
pathologies that arise when the perturbation is performed upon the
Hartree-Fock solution. In the non-equilibrium case these pathologies
are reflected in the unphysical behavior of the dot level charge as a
function of $\epsilon_0$ (around $\epsilon_0 \sim -U/2$); the currents
$I_L$ and $I_R$ exhibit also unphysical features \cite{Hersh}
including violation of current conservation ($I_L \ne I_R$).
A detailed comparison of the results given by the different approaches,
together with a more comprehensive discussion of our method will be
presented in a future publication.

In order to illustrate the kind of results obtained by our method we
have considered two different situations, with the dot level
below (Fig. 2) and above (Fig. 3) the Fermi energy respectively.
In the second case the dot level is almost empty at zero bias.

Figure 2 shows the current intensity, $I$, and the differential
conductance $g = \partial I/\partial V$, as a function of the applied
bias, for $\epsilon_0 = -\pi$ and $U = 0$, $\pi/2$, $\pi$, $2\pi$,
and $4\pi$, where all the energies hereafter are measured in units
of $\Gamma$. Our results for
$g$ show a single broad peak around $V =
2\pi$ for $U = 0$; this corresponds to the dot level
$\epsilon_0 = -\pi$ crossing the right reservoir chemical
potential $\mu^R = -V/2$.
For small $U$ ($U < |\epsilon_0|$) this one-electron like resonance is
shifted and adopts a somewhat asymmetric shape. On the other hand, for
large $U$ (see case $U = 4 \pi$ in Fig. 2) three different features
are clearly present in the current-voltage characteristic. The
conductance peak at $V = 0$ is related to the Kondo resonance
appearing in the dot spectral density around the Fermi energy \cite{Hersh},
while the peaks at $V = 2\pi$ and $V = 4\pi$ correspond to
the crossing of the dot ``ionization'' and ``affinity'' levels, $\epsilon_0$
and $\epsilon_0 + U$, with the reservoirs chemical potentials.
We should comment that these two peaks reflect the charging effects
associated with the dot level: the second level can only be filled
when the applied bias overcomes the repulsion between the second
and the first electron.
For the particular case $U = 2 \pi$ we have a symmetric problem with the
two dot levels crossing the left and right chemical potentials at the
same bias, $V = 2 \pi$,  leading to a single ``charging effect'' peak.
Notice that in this case the conductance at $V = 0$ reaches its maximum
value $2 e^2/h$.
Finally, the case $U = \pi$ illustrates the transition between small
and large $U$, with a new correlation structure arising due to the
overlap between both Kondo and charging effects,
the resonances merging into a broad peak around $V = 0$.
This transition is also apparent in the
current intensity $I(V)$ shown in Fig. 2, where the case $U = \pi$
defines the border between the highly correlated limit (Kondo and
charging effects completely separated) and the one-electron like
behavior.

Figure 3 shows $I(V)$ and $g$ as a function of $V$
for $\epsilon_0 = 3\pi$ and $U = 0$, $2\pi$, $4\pi$ and $6\pi$.
In these cases the dot level is above the Fermi energy, and for $V =
0$ no Kondo-like peak appears in the spectral density of states.
As before, the two peaks appearing in the differential conductance for
$U \ge 2\pi$
are related to the filling of the ionization, $\epsilon_0$, and affinity,
$\epsilon_0 + U$, levels as a function of the applied voltage. As shown
in figure 3b, the voltage difference between the two peaks for $U \ge
2\pi$ is approximately equal to $2U$.
On the other hand, the conductance peak at $V = 2\epsilon_0$
is reduced to nearly half its $U = 0$ value when the interaction increases.
It is also worth noticing that
the $I - V$ curve in this case exhibit a step-like behavior which
is more pronounced for increasing values of $U$. The relative height
of these steps is in agreement with calculations  based on a simple
atomic-like model for the dot Green functions \cite{Chen&Ting}.

Figure 2 and 3 show the different kind of current-voltage
characteristics one gets for the
single dot level described by hamiltonian (1).
For $U$ sufficiently large, two resonances associated with the ionization
and affinity levels can always be observed in the differential conductance.
In this case correlation effects are very important giving rise to an
additional resonance at $V = 0$ when the ionization level is
initially filled.

The results found in this paper suggest that the Kondo-like resonance
and the peaks associated with the charging effects of the quantum level
could be found simultaneously in a quantum dot, provided that the
temperature is sufficiently below the Kondo temperature. As an example, we
consider a $GaAs$-dot of size $L \sim 100 \AA$ both in the vertical
and lateral directions,
sandwiched between two $GaAs$-wires and two
$AlGaAs$-barriers, as those studied in reference \cite{Reed}.
A single bound level around $50 meV$ is found by
solving numerically a simple double barrier model with a barrier
height $\sim$ 300 $meV$. The parameter $U$ can be
evaluated as the Coulomb integral for
the wave function corresponding to this bound level,
which roughly yields $U \sim e^2/\epsilon L \simeq 15 meV$, in agreement
with the dot classical capacitance energy \cite{Chen&Ting}.
Then, varying the $AsGa$-wires doping around $10^{18}/cm^3$, one
could get an experimental device close to
some of the theoretical cases analyzed in figures 2 and 3. In
particular, for $n = 10^{18}/cm^3$ and a barrier width of 30 $\AA$,
we find $\epsilon_0/\Gamma \simeq -3$
and $U/\Gamma \simeq 8$, not far from one of the cases presented in figure 2.
We should recall that finite conduction band effects are not included in
our present calculation; one should always keep in mind that for $V$
sufficiently large the effect of the bottom edge of the semiconducting
wires would be present in the $I-V$ characteristic.

In conclusion, we have presented an accurate solution for the many-body
problem of a single dot level between two biased reservoirs. Our results
show the experimental conditions one should achieve to observe in the
differential conductance a peak related to the Kondo-like structure in
the density of states, and a complementary structure associated with the
charging effects of the dot level. By adjusting appropriately the
dimensions and the doping of a quantum semiconducting dot, we have
shown how
the different peaks could be determined by measuring the differential
conductance.

Support by Spanish CICYT (contract no. PB89-0165) is acknowledged.
We also thank Dr. J. Ferrer, Prof. J.P. Hernandez and Prof. S.Y. Wu for
interesting discussions.

\begin{figure}
\caption{Fulfillment of the Friedel-Langreth sum rule
as a function of the dot level position using the second
order self-energy as calculated: (a) starting form the Hartree-Fock solution
and (b) imposing self-consistency in the dot level charge. The full line
corresponds to the dot level charge $<n_{0 \sigma}>$
and the dotted line corresponds to $-\frac{1}{\pi}\eta(0)$. Inset: second
order diagram used to calculate the self-energies.}
\end{figure}

\begin{figure}
\caption{Current $I$ and differential conductance $g$ obtained using our
method for $\epsilon_0 = -\pi$ and $U$ = 0 (a), $\pi/2$ (b), $\pi$ (c),
$2\pi$ (d), and $4 \pi$ (e).
All the energies are measured in units of the elastic decay rate
at the dot $\Gamma$.}
\end{figure}

\begin{figure}
\caption{Same as figure 2 for $\epsilon_0 = 3\pi$ and $U$ = 0 (a),
$2\pi$ (b), $4\pi$ (c), and $6\pi$ (d).}
\end{figure}
\end{document}